\documentclass[notitlepage,superscriptaddress,showpacs,aps,prd]{revtex4-1}

\usepackage{amsmath}
\usepackage{amsfonts}
\usepackage{amsthm}
\usepackage{amssymb}
\usepackage{graphicx}
\usepackage{epsfig}
\usepackage{hyperref}
\usepackage{amscd}
\usepackage{color}

\begin{document}
\title{ Newman-Janis Ansatz in conformastatic  spacetimes}

\author{Antonio C. Guti\'errez-Pi\~{n}eres}
\email{gutierrezpac@gmail.com}
\affiliation{Instituto de Ciencias Nucleares, Universidad Nacional Aut\'onoma de M\'exico,
 \\AP 70543,  M\'exico, DF 04510, M\'exico}
\affiliation{Facultad de Ciencias B\'asicas,\\
Universidad Tecnol\'ogica de Bol\'ivar, CP 131001,  Cartagena, Colombia}

\author{Hernando Quevedo}
\email{quevedo@nucleares.unam.mx}
\affiliation{Instituto de Ciencias Nucleares, Universidad Nacional Aut\'onoma de M\'exico,
 \\AP 70543,  M\'exico, DF 04510, M\'exico}

 \begin{abstract} 
The Newman-Janis Ansatz was used first to obtain the stationary Kerr metric from the static Schwarzschild metric. Many works have been devoted to investigate 
the physical significance of this Ansatz, but no definite answer has been given so far. We show that this Ansatz can be applied in general to conformastatic 
vacuum metrics, and leads to stationary generalizations which, however, do not preserve the conformal symmetry. 
We investigate also the particular case when the seed solution is given by the Schwarzschild spacetime
and show that the resulting rotating configuration does not correspond to a vacuum solution, even in the limiting case of slow rotation. 
In fact, it describes in general a relativistic fluid with anisotropic pressure and heat flux. 
This implies that the Newman-Janis Ansatz strongly depends on the choice of representation for the seed solution. We interpret this result as 
as  a further indication of its applicability limitations.
\end{abstract}


\maketitle

\section{Introduction}
\label{sec:int}

Stationary solutions in general relativity are very important in the context of relativistic astrophysics.  If we assume axial symmetry in vacuum, the Kerr solution \cite{kerr63} describes the exterior gravitational field of a rotating stationary configuration. A major open problem in classical general relativity is to find
an exact interior solution that could be matched with the exterior Kerr geometry.  
Soon after the discovery of the Kerr solution, Newman and Janis \cite{nja65} showed an
algorithm for obtaining the Kerr solution from the  Schwarzschild solution. The Newman-Janis Ansatz (NJA) can be interpreted as a complex coordinate transformation 
that acts on the Schwarzschild metric for deriving the Kerr solution.
The same method has been used to obtain a Kerr-NUT
solution \cite{new65} and a solution of the Einstein-Maxwell equations \cite{dn66},  starting
from the Schwarzschild and the Reissner-Nordstr\"om metric, respectively.
The NJA was investigated in general  by Talbot \cite{tal69} who proposed the first explanation for its success. 
Demianski \cite{dem72} proved that the Taub-NUT metric with cosmological constant is
the most general solution of Einstein's equations with cosmological constant
that  can be generated by using the NJA.  Of course, the reasons why such a method does work can be traced to the behavior of 
Einstein equations \cite{sam73,fin75}. Moreover, at the level of the curvature tensor it is also possible to apply complex transformations to generate 
new solutions \cite{quev92a,quev92b}.

The NJA was generalized by Herrera and Jim\'enez \cite{herjim82} to include the case of static spherically symmetric interior solutions in order to generate stationary interior spacetimes. Several interior Kerr solutions have been obtained by employing this method. For example, in \cite{dratur97} an interior trial solution was obtained which is characterized by a pressure that diverges at the origin of coordinates. In \cite{ibo05}, several rotating neutral and charged solutions were obtained, describing the interior field of non-perfect fluids. The case of rotating spacetimes for anisotropic fluids with shear viscosity and heat flux was analyzed in detail in
\cite{pap05,pap09}, obtaining some particular solutions whose exterior counterpart is unknown, however. In \cite{via06}, the extension of the NJA was applied to static spacetimes, like the incompressible Schwarzschild interior. The same method can be applied to obtain interior metrics which match a general stationary vacuum spacetime, provided the starting static metric is physically reasonable. Moreover, in \cite{via10}, the field equations for anisotropic fluids were presented in an Ernst-like form which leads to a precise method for generating interior solutions. It turns out that, when applied to static spherically symmetric interior solutions, the extension of the NJA always destroys the perfect-fluid property \cite{ros99}; nevertheless, in the case of a pure gravitomagnetic Weyl tensor, the perfect-fluid property can be preserved \cite{lozwyl11}. 
Although it has been argued \cite{fer14} that the NJA would work only in Einstein's theory. Moreover, the application of the NJA to spherically symmetric solutions of alternative gravity theories  has been shown to lead to pathologies in the resulting axially symmetric spacetimes \cite{hy14}. Nevertheless, 
 it has found applications to  obtain rotating  higher dimensional spacetimes 
\cite{less08}, non-commutative  black holes \cite{modnic10,mxz12}, loop black holes \cite{carmod10}, regular black holes \cite{azr14a,lct15} and  wormholes 
\cite{azr14b}.  Moreover, a generalization of the NJA was proposed which includes the transformation of a particular gauge field \cite{erb15}. 

The main objective of the present work is to apply the NJA to comformastatic spacetimes and, in particular, to the Schwarzschild exterior metric in isotropic coordinates.
We  follow the original terminology introduced by Synge \cite{synge60}, according to which  stationary spacetimes with a conformally flat space of orbits constitute the conformastationary spacetimes, and conformastatic spacetimes comprises the static subset. Several exact solutions belonging to this class have been derived in a series of recent works \cite{ggq13,glq15,gut15,gc15} which are interpreted as describing the gravitational and magnetic fields of static and rotating thin disks. In this work, we start from a general conformastatic metric from which a stationary metric is obtained whose main physical are also analyzed. 

This paper is organized as follows. In Sec. \ref{sec:vac}, we present the field equations for vacuum conformastatic spacetimes, and derive a particular solution.
In Sec. \ref{sec:nja}, we present all the details of the application of the Newman-Janis algorithm to conformastatic spacetimes. In Sec. \ref{sec:phys}, we show that for the resulting metric to be a solution of Einstein's equations it must correspond to a stationary gravitational field with a perfect-fluid source endowed with an additional electromagnetic field. Finally, Sec. \ref{sec:con} contains the final remarks about our results.

\section{Vacuum conformastatic spacetimes}
\label{sec:vac}

Consider  the  following conformastatic  line  element  in spherical  coordinates	$x^{\alpha}=(t, r, \theta, \varphi)$:
     \begin{align}
		\label{eq:confstat_metric}
      ds^2 =V^2 dt^2 - U^4 (dr^2 + r^2d\theta^2 + r^2 \sin^2\theta d\varphi^2)\ .
     \end{align}
In general, the functions $U$ and $V$ can depend on all spatial coordinates. For the sake of simplicity, however, in this work we focus on the
simple case in which these functions depend on the radial coordinate $r$, only. A straightforward computation leads to the following expressions  
      \begin{align}
        R^1_{{\;\;1}} & = (2 r U_r V_r + r U V_{rr} + 2 UV_r)/(r U^5 V),
      \label{eq:Ricci11}\\
      R^2_{{\;\;2}} & = (4  r U V U_{rr} - 4 r V U_r^2 - 2 r U U_r V_r + r U^2 V_{rr}  +  4 U V U_r)/(r U^6 V),
      \label{eq:Ricci22}\\
      R^3_{{\;\;3}  }&= R^4_{ {\;\;4}} = (2 r U V U_{rr}  + 2 r V U_r^2  + 2 r U V_r U_r + 6U V U_r + U^2 V_r)/ (r U^6 V), \label{eq:Ricci33}
     \end{align}
where $R^{\alpha}_{{\;\;\beta} }$ are  the components  of  the  Ricci  tensor.  
In vacuum,  it is easy to show that the above system reduces to     
     \begin{align}
      \nabla^2 U =0,\quad
      \nabla\cdot(U^2 \nabla V) =0,
      \label{eq:UV}
     \end{align}
where  $\nabla$ is the usual gradient  operator  in spherical   coordinates. Thus, $U$  is  a  harmonic  function. Having $U$, the  second equation  in (\ref{eq:UV})
gives $V$. It  is  easy  to prove that  the  only possible functional  dependence   $V=V[U]$ is  of the 
form $V= -kU^{-1}$ with $k=$constant. However, with  this  kind   of  relationship  between $U$ and $V$  the  only  possible  solution for  the complete system 
$R^{\alpha}_{\;\;\beta}=0$  is the  trivial  solution. So, in general, the functions $U$ and $V$ are not related. 

Is not easy to find functions $U$ and $V$  satisfying the  above  system.  However,  the  following  solution does exist
    \begin{align}
    U=c+\frac{b}{r}  ,\quad V=\frac{c r - b}{c r + b}, \quad (a,b \quad \text {constants}),
    \end{align}
which is equivalent to one of the most important solutions of Einstein's equations, namely, the  exterior  Schwarzschild 
solution in isotropic coordinates  \cite{solutions} 
     \begin{align}
         U =1 + \frac{m}{2 r},\quad
         V=\frac{2 r-m}{2r + m}\ ,
         \label{eq:SchwarzschildSol}
       \end{align}
corresponding to $c=1$ and $   b=m/2$.
			
The NJA  is usually applied to obtain new stationary solutions from vacuum static solutions. We will follow the same idea in this work. Indeed, we will assume that we have two functions $U(r)$ and $V(r)$ that satisfy the vacuum conformastatic field equations (\ref{eq:UV}). We will call this set of functions the seed solution. Our goal is to apply the NJA to a seed solution and to investigate the physical properties of the resulting metrics. In particular, we will be interested in finding explicitly the metric resulting from the Schwarzschild seed solution in order to compare our results with the ones obtained originally by Newman and Janis.

\section{The  Newman-Janis Ansatz}
\label{sec:nja}

In this  section, we apply the NJA to a general conformastatic spacetime given by the line element (\ref{eq:confstat_metric}) with the metric functions
$U$ and $V$ satisfying the vacuum field equations (\ref{eq:UV}). Following the procedure presented originally in \cite{nja65}, we introduce the 
  outgoing  Eddington-Finkelstein coordinates $( u, r^*)$ by means of   
     \begin{align}
      u= t-r^*,  \quad d r^*=\frac{U^2}{V} d r \ .
     \end{align}
	Then, the line element 	 (\ref{eq:confstat_metric})  can be written as 
	     \begin{align}\label{eq:confstat_metric_uv}
      ds^2=V^2 du^2  + 2 U^2 V du\, dr - r^2 U^4(dr^2 + r^2 d\theta^2 +  r^2 \sin^2\theta d\varphi^2).
     \end{align}
To apply the NJA to this line element, we introduce the complex null tetrad 
    \begin{align}
      g^{\alpha \beta}  & = l^{\alpha}n^{\beta} + n^{\alpha} l^{\beta} - m^{\alpha}{\bar{m}}^{\beta} -{\bar{m}} ^{\alpha}m ^{\beta},  \nonumber \\
      l^{\alpha} & = \delta^{\alpha}_{r}, \quad
          n^{\alpha}  = \frac{1}{U^2 V} \delta^{\alpha}_u - \frac{1}{2 U^4} \delta^{\alpha}_r, 
           \label{eq:nulltetrad2}\\
           m^{\alpha} & = \frac{1}{\sqrt{2} U^2 r} \delta^{\alpha}_{\theta} - \frac{i}{\sqrt{2} U^2 r \sin{\theta}} \delta^{\alpha}_r, 
           \nonumber
          \end{align}   
where $l^{\alpha}l_{\alpha} = m^{\alpha}m_{\alpha} = n^{\alpha}n_{\alpha} =l^{\alpha}m_{\alpha}=m^{\alpha}n_{\alpha}=0$ and 
$l^{\alpha}n_{\alpha} = -m^{\alpha}{\bar m}_{\alpha} = 1$. Complex conjugation is denoted by a bar over the corresponding quantity. Now,  we  perform the complex transformation
     \begin{align}
      r \rightarrow \tilde{r}   = r  +  i a \cos{\theta},\quad
      u \rightarrow \tilde{u} = u - i a \cos{\theta},\label{eq:ComplexIncrement}\\
      \{U(r), V(r)\} \rightarrow \{M(r,\theta;a), N(r,\theta; a)\} \nonumber
     \end{align}
  where the transformed functions $M$ and $N$ are demanded to be real and to satisfy the condition 
	$\lim_{a \rightarrow 0}\{M,N\}=\{U,V\}$. Thus, the null tetrad (\ref{eq:nulltetrad2}) transforms  into 
         \begin{align}
          l^{\alpha}     & = \delta^{\alpha}_{r},          \quad
          n^{\alpha}    = \frac{1}{M^2 N} \delta^{\alpha}_u - \frac{1}{2 M^4} \delta^{\alpha}_r,    \label{eq:NullTetradTrans2}\\
           m^{\alpha}  & = \frac{(r -  i a \cos{\theta})}{\sqrt{2} M^2 (r^2 +  a^2 \cos^2{\theta})} \delta^{\alpha}_{\theta}   
                                  +\frac{a \sin{\theta (i r +  a \cos{\theta})}}{ \sqrt{2} M^2 (r^2 +  a^2 \cos^2{\theta})}  ( \delta^{\alpha}_u - \delta^{\alpha}_r)
                                  +  \frac{(i r +  a \cos{\theta})}{\sqrt{2} M^2 r \sin{\theta} (r^2 +  a^2 \cos^2{\theta})} \delta^{\alpha}_{\varphi} .
           \nonumber
          \end{align}
From here, we obtain the  transformed  inverse  metric
        \begin{align}
        g^{uu} &= - \frac{a^2 \sin^2{\theta}}{M^4 (r^2 + a^2 \cos^2\theta)},  
         \quad   g^{ur}= \frac{1}{M^2 N} + \frac{a^2 \sin^2{\theta}}{M^4 (r^2 +  a^2 \cos^2{\theta}) }, 
         \quad     g^{u\varphi} = -\frac{a}{ M^4 (r^2 +  a^2 \cos^2{\theta}) },\nonumber\\
         g^{rr}& = -\frac{1}{M^4} - \frac{a^2 \sin^2{\theta}}{M^4 (r^2 +  a^2 \cos^2{\theta})}, 
         \quad    g^{r\varphi}=  \frac{a}{M^4 (r^2 +  a^2 \cos^2{\theta}) },\quad
         g^{\theta\theta} = -\frac{1}{M^4 (r^2 +  a^2 \cos^2{\theta})} \nonumber\\
         g^{\varphi \varphi}&= - \frac{a}{M^4 (r^2 +  a^2 \cos^2{\theta})  \sin^2{\theta}},\nonumber
        \end{align}
and the corresponding line element in Eddington-Finkelstein coordinates 
     \begin{align}\label{eq:Confstat_Metric_EFC}
      ds^2 &  = N^2 du^2 + 2M^2 N du \,dr +  2 a  (M^2 N - N^2) \sin^2{\theta} du \, d\varphi - 2  a M^2 N\sin^2{\theta} dr \,d\varphi - M^4 (r^2 +  a^2 \cos^2{\theta}) d\theta^2\nonumber\\
               &  -  \big[     a^2 (  2 M^2 N - N^2) \sin^2{\theta} + M^4 (r^2 +  a^2 \cos^2{\theta})    \big] \sin^2{\theta} d\varphi^2.
     \end{align}

Now, if we choose the transformed functions as 
     \begin{align}
     M(r, \theta) = U(r)  \quad \text{and} \quad N(r,\theta) = \frac{U^2 (r^2 + a^2\cos^2\theta)}{\frac{r^2 U^2}{V} + a^2\cos^2\theta}
     \end{align}
and introduce  Boyer-Lindquist (BL) like coordinates by means of the coordinate transformation  
     \begin{align}
       du= dt - \frac{\frac{r^2U^2}{V} + a^2 }{r^2 + a^2} dr, \quad d\varphi =d\phi -  \frac{a  }{ r^2  + a^2} dr \ ,\label{eq: CoordinTransformation3}
     \end{align}
the  metric generated by the NJA reduces to   
       \begin{align}
       ds^2 & =  \frac{U^4 (r^2 + a^2 \cos^2\theta)^2}{ (a^2\cos^2\theta + \frac{r^2 U^2}{V} )^2} dt^2 
                   +   \frac{  2 a U^4 (r^2 + a^2 \cos^2\theta)  (\frac{r^2 U^2}{V} - r^2  ) \sin^2\theta }{ (a^2\cos^2\theta + \frac{r^2 U^2}{V} )^2}   dt \;d\phi \nonumber\\
                  &  - \frac{U^4 (r^2 + a^2 \cos^2\theta)}{a^2 + r^2} dr^2  - U^4 (r^2 + a^2 \cos^2\theta) d\theta^2\nonumber\\
                  & -U^4 (r^2 + a^2 \cos^2\theta)\sin^2\theta \left[ 1 + a^2\sin^2\theta \frac{ ( \frac{2 r^2 U^2}{V} - r^2 + a^2\cos^2\theta) }
                       {(a^2 \cos^2\theta + \frac{r^2 U^2}{V}  )^2}  \right] d\phi^2, \label{eq:rotmetric}
       \end{align}
       where,  as  we  mentioned above,  $U(r)$  and $V(r)$  are  a solution of  the   Einstein vacuum equations for  a conformastatic spacetime. 
	Notice  that the above metric can also be obtained as a particular case from Eq.(\ref{eq:rotmetric}) of Ref. \cite{azr14a} 
	where a general static metric is analyzed. To this end, it is necessary to choose the metric functions as 
			    \begin{align}
           G(r) = V^2(r), \quad  H(r)=\frac{r^2}{F(r)},\quad  F(r) = \frac{1}{U^4(r)}, \quad K=\frac{r^2U^2}{V},\quad \psi= U^4 (r^2 + a^2 \cos^2\theta) \ .
   \end{align}
 For our purposes, however, we rewrite the new metric (\ref{eq:rotmetric}) as 
\begin{align}
ds^2  = \frac{V^2\rho^2}{\Sigma^2}\left[ dt^2 - 2 a  \frac{r^2}{\rho} \left(1-\frac{U^2}{V}\right) \sin^2\theta dtd\phi 
+ a^2\left(1-\frac{2\Sigma U^2}{\rho V} \right) \sin^4\theta d\phi^2 \right] 
 -U^4 \rho \left[\frac{dr^2}{r^2+a^2} + d\theta^2 + \sin^2\theta d\phi^2\right] \ ,
\label{gen}
\end{align}
with
\begin{equation}
\rho= r^2 + a^2\cos^2\theta\ , \qquad \Sigma = r^2 +a^2 \cos^2\theta  \frac{V}{U^2} \ .
\end{equation}

Notice that in the limit $a\rightarrow 0$, the generalized metric reduces to the seed metric (\ref{eq:confstat_metric}). Therefore, the parameter $a$ can be interpreted 
as responsible for the stationarity of the spacetime and, consequently, should be associated with the rotation of the gravity source. 
   
\section{Physical interpretation}
\label{sec:phys} 

The generalized metric (\ref{gen}) is clearly stationary; however, it is not conformastationary. This implies that the NJA does not preserve the conformal invariance in this case.  Moreover, if we impose the vacuum field equations for the functions (\ref{eq:UV}) $U$ and $V$, one can show that the corresponding Einstein tensor does not vanish; instead, we compute the following structure 
       \begin{align}
     G_{\alpha\beta} = 
             \begin{pmatrix}       G_{tt}             &           0                    &          0                                    &      G_{t\phi}   \\
                                                0                &      G_{r r}                 &       G_{r\theta}                        &         0         \\
                                                0                &      G_{r\theta }         &       G_{\theta\theta}                &         0         \\
                                           G_{t\phi}    &            0                   &           0                                    &     G_{\phi\phi}         
             \end{pmatrix}. 
             \label{eq:ET} 
       \end{align}     
It follows that in the case of vacuum comformastatic spacetimes the NJA leads in general to non-vacuum stationary spacetimes. 

To investigate the physical interpretation 
of the spacetimes generated in this manner, we proceed as it is customary in the case of conformastatic and conformastic metrics, namely, we search for the conditions under which the above Einstein tensor can be interpreted as corresponding to a perfect-fluid source possibly endowed with an electromagnetic field \cite{ggq13}
in such a way that it could describe the field of a disk-halo configuration.   
   
\subsection{The linearized limit}
\label{sec:srl}

For the sake of simplicity, we consider first the approximate linearized limit in which quadratic terms in $a$ can be neglected. Since the parameter $a$ is related to the rotation of the source, we can expect that in the limiting case of slow rotation the resulting solution is related to the Lense-Thirring solution \cite{solutions}
\begin{equation}
ds^2 = \left(1-\frac{2m}{r}\right) dt^2 - 4 m a \frac{\sin^2\theta}{r} dt d\phi - \frac{dr^2}{1-\frac{2m}{r}} - r^2(d\theta^2 + \sin^2\theta d\phi^2) \ .
\end{equation}
The non-diagonal term corresponds to the gravitational field generated by the rotation of the source. 

The generalized metric (\ref{gen}) reduces in this case to 
\begin{equation}
ds^2 = V^2 \left[ dt^2 - 2 a \left(1-\frac{U^2}{V}\right)\sin^2\theta dtd \phi\right] - U^4\left(dr^2 + r^2d\theta^2 + r^2\sin^2\theta d\phi^2\right) \ .
\label{approx}
\end{equation}
If the metric functions satisfy the conditions $\lim_{r\to\infty} U =\pm 1$ and  $\lim_{r\to\infty} V =1$, the spacetime is asymptotically flat, implying that the source 
of gravity is located in a limited region of spacetime. A direct calculation of the Einstein tensor shows that in general it is non-vanishing.  For the sake of concreteness, let us consider the Schwarzschild metric (\ref{eq:SchwarzschildSol}) as the seed solution. Then, the only non-vanishing component of the Einstein tensor is 
\begin{align}
G_{t\phi}=  32\,{\frac {a m^2 r ( 2r^2 + 3mr - m^2) \sin^2  \theta  }{
 \left( 2\,r+m \right) ^{6} \left( 2\,r -m  \right) }} \ ,
\label{lin}
\end{align}
Remarkably, all the component $G_{tt}$  vanishes in this limit. This means that there is no energy density to be interpreted as the source of gravity. This makes difficult the interpretation of this approximate solution. In fact, one can try to identify the component $G_{t\phi}$ as due to a particular magnetic distribution in Einstein-Maxwell theory, i.e., satisfying the equations
 \begin{align}
      G_{\alpha\beta} =\frac{1}{4 \pi} \big( 
                                 F_{\alpha\mu} F_{\beta \nu} g^{\mu \nu}
                                - \frac{1}{4} g_{\alpha\beta} g^{\mu\lambda} g^{\delta\nu} F_{\mu\delta}F_{\lambda\nu} 
                                  \big) \ ,
        \label{einmax}
       \end{align}           
where $F_{\alpha\beta}$ is the Faraday tensor which can be expressed in terms of the electromagnetic potential $A_\alpha$ as 
$F_{\alpha\beta} = A_{\alpha,\beta} - A_{\beta\alpha}$. One can prove that starting from a general magnetic potential $A_\phi(r)$ 
there is no real solution for the components of the Faraday tensor such that the approximate Einstein-Maxwell equations (\ref{einmax}) are satisfied. 

If the parameter $a$, induced by the NJA, would be related only to the stationary rotation of the gravitational source, we would obtain in the linearized limit the 
Lense-Thirring metric in vacuum or its magnetic generalization. The results presented above show that this is not the case. We can nevertheless force the equivalence 
for particular cases. Indeed, we see that at the pole  $\theta=0$, the Einstein tensor component (\ref{lin}) vanishes. However, along the poles the Lense-Thirring metric predicts no gravitational influence due to  the rotation and, consequently, this particular value corresponds to the original Schwarzschild metric. 
The additional 
particular case for which  $ 2r^2 + 3mr - m^2 =0$ leads to a radius value located inside the horizon which is, therefore, unphysical because it cannot be detected by an 
observer located outside the horizon. 
Thus, we see that it is not possible to recover the
Lense-Thirring metric in any particular case.

\subsection{A relativistic fluid}
\label{CanonicalEMT}

The non-diagonal structure of the Einstein tensor (\ref{eq:ET}) for the general stationary metric indicates that it cannot be interpreted as 
a perfect fluid. Moreover, a straightforward calculation of its trace shows that it is in general different from zero, indicating that 
the identification with the electromagnetic Maxwell tensor is possible only in very special cases.  
Let us therefore consider the general case of a relativistic fluid whose energy-momentum tensor is given by
\begin{align}
               T_{\alpha\beta}&= (\mu + P)V_{\alpha}V_{\beta} - P g_{\alpha\beta}
                               + {\cal Q}_{\alpha}V_{\beta} + {\cal Q}_{\beta}V_{\alpha}
                               + \Pi_{\alpha\beta}\label{eq:EMTcanonical},
                 \end{align}
where  $\mu$ and $P$ represent the energy and pressure, respectively, the worldlines of the fluid are integral curves of the 4-velocity vector field 
$V^\alpha$, the heat flux vector is ${\cal Q}_{\beta}$, and $ \Pi_{\alpha\beta}$ represents the viscosity tensor. Notice that  ${\cal Q}_{\beta}$, and 
$ \Pi_{\alpha\beta}$ are transverse to the worldlines of the fluid in the sense that ${\cal Q}_{\alpha}V^{\alpha}={\cal Q}^{\alpha}V_{\alpha}=0$, 
and  $ \Pi_{\alpha\beta} V^\alpha=0$. If a particular solution of Einstein's equations is described by  the energy-momentum  tensor (\ref{eq:EMTcanonical}),
we may say that the gravitational field is generated by a source in which  $\mu$, $P$, ${\cal Q}_{\alpha}$ and $\Pi_{\alpha\beta}$ are 
 the  energy density, the isotropic  pressure,  the  heat flux and the anisotropic  tensor of  the source.  
Thus, it  is  straightforward to see  that 
                              \begin{align}
                                   \mu &=T_{\alpha\beta} V^{\alpha}V^{\beta},\label{eq:energy}\\
                                      P&= \frac{1}{3}  {\cal H}^{\alpha\beta}T_{\alpha\beta},\label{eq:pressure} \\
                      {\cal Q}_{\alpha}&= T_{\alpha\beta} V^{\beta}  - \mu V_{\alpha} ,\label{eq:heatflux}\\
                      \Pi_{\alpha\beta}&={\cal H}_{\alpha}^{\;\;\mu}{\cal H}_{\beta}^{\;\;\nu}
                                      ( T_{\mu\nu}  -  P{\cal H}_{\mu\nu}), \label{eq:anisotropict}
                                          \end{align}\                            
where  the  projection  tensor  is  defined  by ${\cal H}_{\alpha\beta} \equiv V_\alpha V_\beta - g_{\alpha\beta}$.

In order  to interpret  the   solution generated by the NJA,   we  first  rewrite  the  metric (\ref{eq:rotmetric})  as
          \begin{align}
            ds^2= A dt^2 + 2 B dt d\phi - \frac{\psi}{a^2 + r^2} dr^2 - \psi d\theta^2 - C d {\phi}^2 \ ,
           \end{align}
where
     \begin{align*}
       A & \equiv \frac{\psi (r^2 + a^2 \cos^2{\theta})}{(K + a^2 \cos^2{\theta})^2} \ , \quad
       B \equiv  \frac{a \psi \sin^2{\theta} (K - r^2)}{(K + a^2 \cos^2{\theta})^2} \ ,  \quad
       C \equiv \frac{\psi \sin^2{\theta} \big[  (K + a^2)^2 - a^2 \sin^2{\theta} (r^2+ a^2)  \big]}
                     {(K + a^2 \cos^2{\theta})^2},\\
                     & \\
       \psi  & \equiv U^4 (r^2 + a^2 \cos^2{\theta}) \ ,  \quad K \equiv  \frac{r^2 U^2}{V}.
     \end{align*}  

It is convenient to introduce a  suitable reference  frame in terms  of  an orthonormal tetrad for a local observer in the  form
       \begin{align}
        V_{\alpha}   & =  \left\{    \sqrt{A} ,  \; 0 ,    \; 0 ,    \;  \frac{B}{\sqrt{A}}  \right\},\\
        K_{\alpha}   & =  \left\{    0 ,          \;    - \frac{\sqrt{\psi}}{ \sqrt{a^2 + r^2}},    \;     0 ,      \;     0 \right\},\\
        L_{\alpha}    & =  \left\{    0 ,          \;       0 \;  - \sqrt{\psi},   \;    0,    \;     0 \right\},\\
        M_{\alpha}   & =  \left\{    0 ,          \;     0 \;    \;    0,    \;  -\frac{ \sqrt{B^2 + A C} }{ \sqrt{A} } \right\}.
            \end{align}
         
         with  the  corresponding  dual  tetrad   
            
                \begin{align}
        V^{\alpha}    &  =  \left\{    \frac{1}{\sqrt{A}} ,  \; 0 ,    \; 0 ,    \; 0 \right\},\\
        K^{\alpha}   &  =  \left\{    0 ,          \;     \frac{ \sqrt{a^2 + r^2}}{\sqrt{\psi}},    \;     0 ,      \;     0 \right\},\\
        L^{\alpha}    &  =  \left\{    0 ,          \;       0 ,  \;    \frac{1}{\sqrt{\psi}} ,    \;     0 \right\},\\
        M^{\alpha}  &  =  \left\{ - \frac{B}{\sqrt{A(B^2 + A C)}} ,          \;     0 \;    \;    0,    \;   \frac{\sqrt{A} }{ \sqrt{B^2 + A C} } \right\}.
            \end{align}
It is easy  to  see that $V^\alpha V_\alpha = - K^\alpha K_\alpha = -L^\alpha L_\alpha = -M ^\alpha M_\alpha  = 1 $    and that
      $V^\alpha K_\alpha =  V^\alpha L_\alpha = V^\alpha M_\alpha = K ^\alpha L_\alpha  = K ^\alpha M_\alpha =  L ^\alpha M_\alpha =0 $.  

The idea now is to verify if Einstein's equations, $G_{\alpha\beta} = 8\pi T_{\alpha\beta}$, which for the particular Einstein tensor (\ref{eq:ET}) and the energy-momentum tensor 
(\ref{eq:EMTcanonical}) lead to an algebraic system of equations, can be solved in a consistent manner and without imposing additional conditions on the components of 
$G_{\alpha\beta}$. To this end, 
we use the constitutive equations  (\ref{eq:energy}) - (\ref{eq:anisotropict})  and  Eq.(\ref{eq:ET}). First,               we can write  the   energy density   and  the  pressure  of  the fluid as 
         \begin{align}
            \mu = \frac{G_{tt}}{A}
          \end{align}
and
               \begin{align}
                        P=\frac{G_{tt} - A G}{ 3 A},
                            \end{align}
respectively, where $G = G_{\alpha}^{\; \; \alpha}$.   These quantities must satisfy the corresponding  energy conditions to be physically meaningful. Furthermore, 
the  heat   function   is  given by
                 \begin{align}
                  {\cal Q}_{\alpha}=    \frac{ (A G_{t \phi} - B G_{tt} ) \delta_{\alpha}^{\;\phi} }{A^{3/2}} \ ,
                    \end{align}              
indicating the heat flux occurs only along the azimuthal direction. Finally, the  non-zero  components  of  the  anisotropic tensor  are
                 \begin{align}
                 \Pi_{rr}& =  G_{rr}  - \frac{\psi}{3 (a^2 + r^2) A}  G_{tt} + \frac{\psi }{3 (a^2 + r^2)} G, \\
                 \Pi_{\theta\theta}& =  G_{\theta\theta}  - \frac{\psi }{3  A} G_{tt} + \frac{\psi }{3} G, \\
                 \Pi_{ r \theta}& =  G_{r \theta}, \\
                 \Pi_{\phi\phi}& =  G_{\phi\phi}  +  \frac{2 B^2 - AC}{3 A^2} G_{tt} - \frac{2 B}{A}G_{t\phi} +  \frac{B^2 + AC}{3 A} G, \\
                     \end{align}          
 In terms of the components of the local orthonormal tetrad, the anisotropic tensor can be decomposed as 
            \begin{align}
             \Pi_{\alpha\beta} =     P_{r} K_{\alpha} K_{\beta} 
                                                                +  P_{\theta} L_{\alpha} L_{\beta}
                                                                +  P_{\phi} M_{\alpha} M_{\beta} 
                                                                +  P_{T}  ( L_{\alpha} K_{\beta}   +  L_{\beta} K_{\alpha} ) \ ,
                    \end{align}
where 
                \begin{align*}
                        P_{r}         &  =  \frac{a^2 + r^2}{\psi} G_{rr} - \frac{1}{3 A} G_{tt} +  \frac{1}{3} G,\\
                       P_{\theta}  &  = \frac{1}{\psi} G_{\theta\theta} - \frac{1}{3 A} G_{tt} +  \frac{1}{3} G,\\
                       P_{\phi} & =  \frac{A}{B^2 + A C } G_{\phi\phi} + \frac{2 B^2 - A C}{3 A (B^2 + AC)} G_{tt}
                                              - \frac{2 B}{B^2 + A C} G_{t\phi} + \frac{1}{3} G,\\
                      P_{T}         &  =   \frac{\sqrt{a^2 + r^2}}{\psi},
                                           \end{align*}
represent the values of the anisotropic pressure in different spatial directions. 
Notice that  ${\cal P} \equiv P_r + P_{\theta} + P_{\phi}=0 $ and, consequently, the trace    $\Pi_{\; \; \alpha}^{ \alpha} =0$.  Notice  also that to get  a fluid  without  heat flux it is  mandatory that  $ A G_{t \phi} - B G_{tt}  = 0$, a condition which in general is not satisfied. Even if we consider the particular metric generated from the Schwarzschild metric in isotropic coordinates, the heat flux cannot be made to vanish.

We see that the NJA generates in this case metrics with properties that are completely different from the properties of the starting seed metric. Consider, for instance,
the Schwarzschild metric as seed solution. The NJA generates a non-vacuum solution for a relativistic fluid in which the heat flux is non-trivial and all the components of the anisotropic pressure are different from zero. It is interesting to consider in this case the limit $m\rightarrow 0$ with $a\neq 0$. The obtained solution has a non-trivial form, but a straightforward computation shows that its curvature tensor vanishes. This an essential logical test which shows that there is no pressure and heat flux without mass. Nevertheless, the resulting solution has a quite complicated physical interpretation which somehow is not exactly the idea of NJA.

		\section{Conclusions}
		\label{sec:con}

The NJA was proposed more that half a century ago as a method to obtain the rotating Kerr solution from the static Schwarzschild solution, but it has also been shown to 
work for the Reissner-Nordstr\"om metric from which the Kerr-Newman solution is generated.  It has been used extensively
to generate stationary perfect-fluid solutions from static ones. Despite its success, the reason why the NJA works is still unknown. In fact, as it is used today, it can be understood as an Ansatz or as a trick. To upgrade the NJA to the status of  an algorithm or a genuine method to generate new solutions of Einstein's equations, it must be described as an exact mathematical formalism that allows us to understand why it can be used to generate new solutions. 
The results obtained in the present work can be interpreted as an indication that the NJA is just a trick that happens to work under very particular circumstances.

In fact, we first applied the NJA to conformastatic vacuum metrics in the hope that we could generate conformastationary vacuum solutions. Our results show that 
this is not possible. Although the resulting metric has a non-diagonal term which is usually associated with the rotation of the source, it does not preserve the
conformal symmetry of the static case. Moreover, the resulting metric does not satisfy Einstein's equations in vacuum. To analyze the physical significance of the 
metric generated by the NJA, we compare its Einstein tensor with the energy-momentum tensor for a relativistic fluid with anisotropic pressure and heat flux. 
We proved that in general all the physical quantities determining the fluid can be identified in a consistent manner with the non-zero components of the Einstein tensor. If we consider the particular case 
of the Schwarzschild metric as seed solution, all the physical quantities of the generated relativistic fluid satisfy the physical condition of vanishing 
as $m\rightarrow 0$, independently of the value of the rotational parameter $a$.

We thus see that the NJA does not generate the vacuum Kerr metric from the Schwarzschild metric in isotropic coordinates. Even in 
the limiting case of small $a$, the resulting linearized solution cannot be identified as the vacuum Lense-Thirring metric. 
This implies that the NJA depends on
the choice of coordinates. We interpret this result as indication that the NJA cannot be considered as an algorithm; instead, it should be interpreted as a trick that 
happens to work well for particular solutions in spherical coordinates. Recently, in \cite{hanyun13}, an additional negative fact about the NJA was observed, namely,
that in the context of modified gravity theories it leads to pathologies in the resulting metrics and that it should not be used to generate rotating black holes outside 
general relativity. In the present work, we show that even within general relativity it should be used with caution to construct black hole solutions, because it depends
on the coordinates used for the construction.

		\section*{Acknowledgments}
		
This work was partially supported by DGAPA-UNAM, Grant No. 113514, and Conacyt, Grant No. 166391.


\begin{thebibliography}{99}

\bibitem{kerr63} R. P. Kerr, Phys. Rev. Lett. {\bf 11}, 237-238 (1963).

\bibitem{nja65} E. T. Newman and A. I. Janis, J. Math. Phys. {\bf 6}, 915–917 (1965).

\bibitem{new65} E. T. Newman  et al.,  J. Math. Phys. {\bf 6}, 918  (1965).

\bibitem{dn66} M. Demianski and E. T. Newman, Bull Acad. Polon. Sci. Set. Math. 
Astron. Phys. {\bf 14}, 653  (1966). 

\bibitem{tal69} C. G.  Talbot, Commun. Math. Phys. {\bf 13}, 45 (1969). 

\bibitem{dem72} M. Demianski,  Phys. Lett. A {\bf 42}, 157 (1972).

\bibitem{sam73} M. Schiffer, R. Adler, J. Mark and C. Sheffield, J. Math. Phys. {\bf 14} , 52–56 (1973).

\bibitem{fin75} R. J. Finkelstein, J. Math. Phys. {\bf 16}, 1271–1277 (1975).

\bibitem{quev92a} H. Quevedo, Gen. Rel. Grav. {\bf 24}, 693 (1992).

\bibitem{quev92b} H. Quevedo, Gen. Rel. Grav. {\bf 24}, 799 (1992).


\bibitem{herjim82} L. Herrera and J. Jim\'enez,
J. Math. Phys. {\bf 23}, 2339 (1982).

\bibitem{dratur97}  S. Drake and  R. Turolla, Class. Quantum. Grav. {\bf 14}, 1883 (1997). 

\bibitem{ibo05} N. Ibohal, Gen. Rel. Grav. {\bf 37}, 19 (2005). 

\bibitem{pap05} T. Papakostas, J. Phys: Conf.Ser. {\bf 8}, 22 (2005). 

\bibitem{pap09} T. Papakostas, J. Phys: Conf.Ser. {\bf 189}, 012027 (2009). 

\bibitem{via06} S. Viaggiu, Int. J. Modern. Phys. D {\bf 15}, 1441 (2006). 

\bibitem{via10} S. Viaggiu, Int. J. Modern. Phys. D {\bf 19}, 1783 (2010). 

\bibitem{ros99} K. Rosquist, Class. Quantum Grav. {\bf 16}, 1755  (1999). 

\bibitem{lozwyl11} C. Lozanovski and L. Wylleman, Class. Quantum. Grav. {\bf 28}, 075015 (2011). 

\bibitem{fer14} R. Ferraro, Gen. Rel. Grav. {\bf 46}, 1705 (2014). 

\bibitem{hy14} D. Hansen and N. Yunes, Phys. Rev. D {\bf 88}, 104020 (2013).


\bibitem{less08} G. Lessner, Gen. Rel. Grav. {\bf 40}, 2177 (2008). 

\bibitem{modnic10} L. Modesto and P. Nicolini, Phys. Rev. D {\bf 82}, 104035 (2010). 

\bibitem{mxz12} Y. Miao, Z. Xue, and S. Zhang, Int. J. Mod. Phys. D {\bf 21}, 1250017 (2012). 

\bibitem{carmod10} F. Caravelli and L. Modesto, Class. Quantum. Grav. {\bf 27}, 245022 (2010). 

\bibitem{azr14a} M. Azrag-Ainou, Eur. Phys. J. C {\bf 74}, 2865 (2014). 

\bibitem{azr14b} M. Azrag-Ainou, Phys. Rev. D {\bf 90}, 064041 (2014). 

\bibitem{lct15} A. Larranaga, A. Cardenas-Avendano, and  D. Torres, Phys. Lett. B {\bf 743}, 492 (2015). 

\bibitem{erb15} H. Erbin, Gen. Rel. Grav. {\bf 47}, 19 (2015).



\bibitem{synge60} J. Synge, Relativity: the general theory, North-Holland Pub. Co. Interscience Publishers, Amsterdam, New York (1960)

\bibitem{ggq13} A. C. Guti\'errez-Pi\~neres, G. A. Gonz\'alez, and H. Quevedo, Phys. Rev. D {\bf 87}, 044010 (2013).

\bibitem{glq15}  A. C. Guti\'errez-Pi\~neres, C. S. Lopez-Monsalvo, and H. Quevedo, Gen. Rel. Grav. {\bf 47}, 1 (2015).

\bibitem{gut15} A. C. Guti\'errez-Pi\~neres, Gen. Rel. Grav. {\bf 47}, 54 (2015). 

\bibitem{gc15} A. C. Guti\'errez-Pi\~neres and A. J. Capistrano, Adv. Math. Phys. {\bf 15}, 2015 (2015).


\bibitem{solutions}  H. Stephani, D. Kramer,
   M. MacCallum, C. Hoenselaers,  and E. Herlt, Exact solutions of Einstein's field equations, Cambridge University Press (2009)

\bibitem{hanyun13} D. Hansen and N. Yunes, Phys. Rev. D {\bf 88}, 104020 (2013).

\end{thebibliography}

\end{document}